\documentclass[10pt,preprint,letterpaper]{article}
\usepackage{ol2}

\usepackage{amsfonts}
\usepackage{amsmath}
\usepackage{amssymb}
\usepackage{graphicx}%
\usepackage{array}
\newcolumntype{I}{!{\vrule width 2pt}}
\newlength\savedwidth
\newcommand\whline{\noalign{\global\savedwidth\arrayrulewidth
                            \global\arrayrulewidth 2pt}%
                   \hline
                   \noalign{\global\arrayrulewidth\savedwidth}}
\newlength\savewidth
\newcommand\shline{\noalign{\global\savewidth\arrayrulewidth
                            \global\arrayrulewidth 1.0pt}%
                   \hline
                   \noalign{\global\arrayrulewidth\savewidth}}

\begin{document}
\twocolumn[
%\renewcommand{\thefootnote}{\scriptsize{\textcircled{\tiny{\arabic{footnote}}}}}
%\title{The Hartman effect in the optical tunneling process beyond the special theory of relativity}
 \title{Enhanced absorption of monolayer MoS$_{2}$  with resonant back reflector}

\maketitle

\author{Jiang-Tao Liu,$^{1,2,*}$  Tong-Biao Wang,$^2$ Xiao-Jing Li,$^{3}$ and Nian-Hua Liu$^{1,2}$}

\address{
$^1$Nanoscale Science and Technology  Laboratory,  Institate for Advanced Study, Nanchang University, Nanchang 330031, China\\
$^2$Department of Physics, Nanchang University, Nanchang 330031, China\\
$^3$College of Physics and Energy, Fujian Normal University, Fuzhou 350007, China\\
$^*$Corresponding author: jtliu@semi.ac.cn}

\begin{abstract}
By extracting the  permittivity of monolayer MoS$_{2}$ from experiments, the  optical absorption of monolayer MoS$_{2}$ prepared on top of one-dimensional photonic crystal (1DPC) or metal films is investigated theoretically. The 1DPC and metal films act  as resonant back reflectors that can enhance  absorption of monolayer MoS$_{2}$ substantially  over a broad spectral range due to the Fabry-Perot cavity effect. The absorption of monolayer MoS$_{2}$ can also be tuned by varying either the distance between the monolayer MoS$_{2}$ and the back reflector or the thickness of the cover layers.
\end{abstract}

%\ocis{(160.6000) Semiconductor materials; (310.6860) Thin films, optical properties;  ( 250.0250) Optoelectronics}

]

Monolayer MoS$_{2}$ as a new kind of  two dimensional (2D) semiconductor has elicited significant attention because of its distinctive electronic and optical properties \cite{12QHW,10KFM,10AS,07TL,13OLS,11ZY,12HSL,12WC,13MB,12JP,14CJ,14AS,13XG,12YVB}.   Monolayer MoS$_{2}$ exhibits  a direct band gap in the visible frequency range \cite{12QHW,10KFM,10AS,07TL},  which is more favorable for optoelectronic applications than graphene in numerous  cases. Monolayer MoS$_{2}$ has show numerous potential applications in flexible phototransistors, photodetectors, photovoltaics,  and signal amplification \cite{13OLS,11ZY,12HSL,12WC,13MB,12JP,14CJ,14AS,13XG}. Notably, the photoresponsivity of monolayer MoS$_{2}$ photodetectors can reach 880 A/W, which is 10$^{6}$ better than that of the first graphene photodetectors ($\sim$0.5 mA/W) \cite{13OLS}.

The optical absorbance in monolayer MoS$_{2}$ is minimal ($<11\%$) due to its ultrathin thickness, which is not conducive to  fabrication of photodetectors, solar cells,   and optical amplification.  Thus, to promote the applications of monolayer MoS$_{2}$, the  optical absorptance in monolayer MoS$_{2}$ waves should be enhanced. In  studies of graphene, several mechanisms have been proposed to enhance the absorption of graphene, e.g.,   periodically patterned graphene, surface plasmon,  microcavity,  graphene-negative permittivity metamaterials, and  attenuated total reflectance, etc \cite{12ST,12AFa,12MAV,12AFb,08ZZZ,13WZ,13QY,12RA,13XZ}. The interaction between graphene and optical beams can also be enhanced when the graphene layers are prepared on top of one-dimensional photonic crystal (1DPC) or with resonant metal back reflectors because of the Fabry-Perot (F-P) cavity effect \cite{12JTL,13NMRP,13JTL}. The proposed structures are very easy to fabricate using existing technology.

In this Letter, the optical absorption of monolayer MoS$_{2}$ prepared on top of 1DPC or metal films with a spacer layer and cover layers is investigated theoretically. We find that the absorption of monolayer MoS$_{2}$ can be enhanced by nearly four times because of the F-P interference. The absorption of monolayer MoS$_{2}$ with 1DPC is slight larger than that of monolayer MoS$_{2}$ with metal films. However, the full width at half maximum (FWHM) of the absorption spectrum  of monolayer MoS$_{2}$ with metal films is much larger than that of monolayer MoS$_{2}$ on top of 1DPC.  The absorption of monolayer MoS$_{2}$ can also be tuned by varying  the thickness of  spacer layers and cover layers. Our proposal is very easy to implement and has potential important applications in monolayer MoS$_{2}$ optoelectronic devices.

The details of the structure are shown in the inset of Fig. 1 (a). The 0.65 nm monolayer MoS$_{2}$ is prepared on top of the SiO$_{2}$ spacer layer with high-$\kappa$ dielectric HfO$_{2}$ cover layers. A 1DPC or 130 nm silver film is placed at the bottom of the SiO$_{2}$ spacer layer as the resonant back reflector (RBR). The  1DPC is composed of alternating MgF$_{2}$ and ZnS layers with a total of 8.5 periods. The permittivity for silver film is frequency dependent \cite{85EDP}. The refractive  indices  of SiO$_{2}$,  HfO$_{2}$, MgF$_{2}$, and ZnS at $\lambda=550$ nm are $n_{SiO_{2}}=1.55$, $n_{HfO_{2}}=1.93$, $n_{MgF_{2}}=1.38$, and $n_{ZnS}=2.59$, respectively \cite{85EDP}.

\begin{figure}[t]
\centering
\includegraphics[width=0.9\columnwidth,clip]{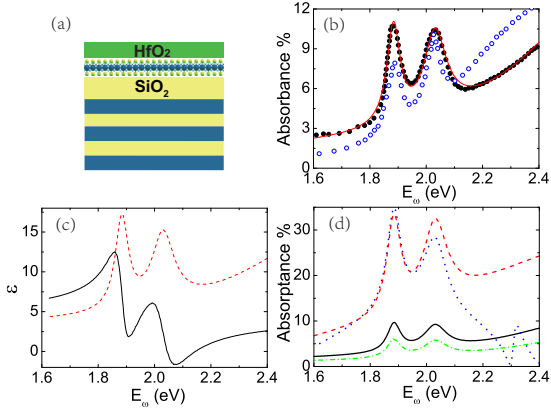}
\caption{(a) Schematic of monolayer MoS$_{2}$ prepared on top of 1DPC with SiO$_{2}$ spacer layer and HfO$_{2}$ cover layer. (b)  The absorbance of monolayer MoS$_{2}$ as a function of the light wavelength of the experimental results (black solid-circle curve),  fit results (red solid line), and the theoretical results in Ref. \cite{13DYQ}  (blue open circle curve). (c) The real part (black solid line) and the imaginary part (red dashed line) of the permittivity of monolayer MoS$_{2}$. (d) The absorptance of monolayer MoS$_{2}$ as a function of the light wavelength without cover layers for different structures: suspended monolayer MoS$_{2}$
(black solid line), monolayer MoS$_{2}$ with the 1DPC and 106 nm spacer layer  (blue dotted line),
monolayer MoS$_{2}$  with sliver films and 75 nm spacer layer (red dashed line), and monolayer MoS$_{2}$ on the on top of thick SiO$_{2}$ substrate (green dash-dotted line).
}
\label{fig1}%
\end{figure}

%The theoretical  results obtained by Qiu et al.  were close to the experimental results. However, the calculations involved are complicated \cite{13DYQ}.  In the calculation, self-energy,  excitons, electron-phonon effects, and ab initio broadening are included. In addition, high cutoff energy $E_{s}=476$ eV, 6000 bands, and $120\times120$ k-points are used, indicating that a large amount of numerical calculations is needed.

To model the absorption of monolayer MoS$_{2}$ in these structures, the details of the  permittivity of monolayer MoS$_{2}$ must be obtained first. Numerous  methods based on first-principle calculations have been used in the calculations of the optical spectrum of monolayer MoS$_{2}$ \cite{13DYQ,13HS,12TC,12AR}. However, the reported theory results are conflicting. Differences between  theoretical  results and  experiment results have been noted.
The permittivity of monolaye MoS$_{2}$ can be extracted from experiments with the use of two exciton transition and the band transition \cite{10KFM}.  The imaginary permittivity of exciton can be described by  Lorentzian function. Neglecting the variation of the transition matrix elements with energy, the band absorption of a 2D semiconductor $\alpha_{b}$ can be described by the step function and the corresponding imaginary permittivity is $\varepsilon_{b}=n_{r0}c_{0}\alpha_{b}/\omega$, where $c_{0}$ is the speed of light in vacuum and $n_{r0}=2.05$ is the transverse component of the static refractive index \cite{12TC}.  Thus, the imaginary part of the total permittivity can be expressed as
\begin{eqnarray}
\varepsilon _{i} &=&\frac{f_{ex}^{A}\Gamma _{A}}{(E_{\omega
}-E_{ex}^{A})^{2}+\Gamma _{A}^{2}}+\frac{f_{ex}^{B}\Gamma _{B}}{%
(E_{\omega }-E_{ex}^{B})^{2}+\Gamma _{B}^{2}} \nonumber \\
&+&\frac{f_{b}e}{\hbar\omega}\Theta (E_{\omega } -E_{g}',\Gamma_{band}), \label{EQ_eps}
\end{eqnarray}
where  $\Gamma _{A}$ ($\Gamma _{B}$),  $f_{ex}^{A}$ ($f_{ex}^{B}$), and $E_{ex}^{A}$ ($E_{ex}^{B}$) are the linewidth, equivalent oscillator strength, and transition energy of  A (B) excitons, respectively,  $E_{g}'$ is the band gap of monolayer MoS$_{2}$,   $f_{b}$ is the  equivalent oscillator strength of interband transition,  $\Theta(x,\Gamma)= \frac{1}{\pi }\int_{-\infty }^{x}\frac{\Gamma }{\Gamma ^{2}+\psi^{2}}d\psi $ is the  step function with a broadening of $\Gamma$, where $\Gamma_{band}$ is the linewidth of the interband transition. The spin-orbit splitting is negligible in the band absorption because spin-orbit splitting is much smaller than the band gap. The real part of the permittivity of monolayer MoS$_{2}$ can be obtained using Kramers-Kronig relations
\begin{equation}
\varepsilon _{r}\left( \omega \right) =n_{r0}^{2}+\frac{1}{\pi }%
\mathfrak{p}\int_{0}^{\infty }\frac{s\varepsilon_{i}\left( \omega \right) }{%
s^{2}-\omega ^{2}}ds, \label{EQ_epsr}
\end{equation}
where $\mathfrak{p}$ is the principal value integral.

The parameters in Eq. (1) are obtained by fitting the experimental data  using the standard transfer matrix method \cite{12JTL}. The results are shown in Table 1 and Fig. 1(b). Although nine fitting parameters are in Eq. (1), these parameters are relatively independent. For instance, the parameters $\Gamma _{A}$,  $f_{ex}^{A}$, and $E_{ex}^{A}$ are decided by the absorption peak located at about 1.88 eV. However, the parameters $\Gamma _{B}$, $f_{ex}^{B}$, and $E_{ex}^{B}$ are obtained by fitting the absorption peak located at 2.02 eV.  The linewidth of the band transition is much larger than that of the traditional semiconductor quantum well, which may have been caused by the strong coulomb interaction and the strong electron-phonon interaction. The folds of  monolayer MoS$_{2}$ and the interaction with the substrate may also enhance the linewidth. The fitting $E_{g}'$ of the absorption spectra is small, which indicates that it is mainly contributed by  the quasiparticle band absorption.

\begin{table}[t]
  \caption{Fitted parameters in Eq. (1)}
  \begin{center}
    \begin{tabular}{ccccccccc}
    \hline
    \hline
    $E_{ex}^{A}$ & $\Gamma _{A}$ & $E_{ex}^{B}$ & $\Gamma _{B}$ & $f_{ex}^{A}$ \\
    \hline
    1.884 eV & 28 meV &  2.02 eV & 42 meV & {\color{red}0.32 eV}\\
    \hline
    \hline
       $f_{ex}^{B}$ & $E_{g}$ &  $\Gamma_{band}$ & $f_{b}$\\
    \hline
      {\color{red} 0.43 eV} & 2.43 eV & 0.398 eV & 59 \\
    \hline
    \hline
    \end{tabular}
  \end{center}
\end{table}

The monolayer MoS$_{2}$ has a large imaginary part of permittivity [Fig. 1(c)]. However, the maximum absorptance (i.e., the directly proportional ratio of the radiation absorbed by the monolayer MoS$_{2}$ to that incident upon it) of suspended monolayer  MoS$_{2}$ is only about 9.6\% due to its ultrathin thickness. By contrast, when a monolayer  MoS$_{2}$ is prepared on top of the 1DPC (silver film), the maximum absorptance can be as high as 34.9\% (33.5\%)[Fig. 1(c)]. Thus, the absorptance of monolayer MoS$_{2}$ with  1DPC (silver) film   can be enhanced by about 3.64 (3.49) times. In this structures,  the monolayer MoS$_{2}$ and the RBR  act as the mirrors of the F-P cavity. The light propagates back and forth in the F-P cavity, which  enhances the absorption of monolayer MoS$_{2}$. The maximum absorption of monolayer  MoS$_{2}$ with 1DPC is slightly larger than that of monolayer  MoS$_{2}$ with a silver film due to the higher reflectivity of  1DPC.  However,  the reflectivity of  1DPC is limited to the photonic band gap width in 1DPC. By contrast, the silver film can reflect lights effectively within the visible light  range.  Thus, the FWHM of the absorption spectrum of monolayer  MoS$_{2}$  with silver film is much larger than that of monolayer  MoS$_{2}$  with 1DPC.  To contrast, we also  show the absorptance of the monolayer  MoS$_{2}$ prepared on top of thick  SiO$_{2}$ substrate in fig. 1(d). The maximum absorptance is only about 6.1\%, which is smaller than that of suspended monolayer  MoS$_{2}$ due to that the traditional substrate material enhance the reflection of the  monolayer  MoS$_{2}$.

\begin{figure}[t]
\centering
\includegraphics[width=0.9\columnwidth,clip]{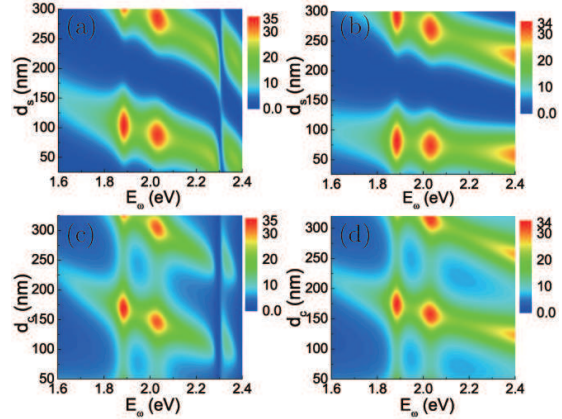}
\caption{Contour plots of the absorptance of the monolayer MoS$_{2}$  with (a) 1DPC and (b) silver films as function of the photon energy and the spacer layers thickness $d_{s}$. Contour plots of the absorptance of the monolayer MoS$_{2}$  with (c) 1DPC and $d_{s}=106$ nm  and (d) silver films and $d_{s}=75$ nm  as function of the photon energy and the cover layers thickness $d_{c}$.
}
\label{fig2}%
\end{figure}

The absorption of monolayer MoS$_{2}$ can also be tuned by varying the  thickness of the spacer layer. In Figs. 2(a) and 2(b), we plot the absorptance of monolayer MoS$_{2}$ as a function of the wavelength and the spacer layers thickness $d_{s}$ without cover layers.  Similar to the microcavity, the resonant absorption wavelength of monolayer MoS$_{2}$ with 1DPC can be described as $n_{Sio_{2}}d_{s}/\lambda=4m\pi$ with including the half-wave loss, where m is an integer. Thus, the absorption peak photon energy increases  with decreasing spacer layer thickness.  However, the resonant absorption wavelength of monolayer MoS$_{2}$ with silver films is not strictly equal to $n_{Sio_{2}}d_{s}/4m\pi$ duo to the dispersion and existing skin  depth in the silver films. The absorption peak photon energy of monolayer MoS$_{2}$ with silver films increases slowly with decreasing spacer layer thickness. The electron mobility in monolayer MoS$_{2}$ can be enhanced several times by the high-$\kappa$ dielectric HfO$_{2}$ cover layers \cite{11BR,11YY}. The cover layers can also adjust the absorption of monolayer MoS$_{2}$. The details are shown in Figs. 2(c) and 2(d). The normal cover layers cannot enhance the absorptance of monolayer MoS$_{2}$. However, the peak photon energy can be tuned by varying the thickness of the cover layers.  Different from the spacer layer, the absorption of monolayer MoS$_{2}$ with both 1DPC  and silver films are the same way when the cover layer thickness increases.

\begin{figure}[t]
\centering
\includegraphics[width=0.9\columnwidth,clip]{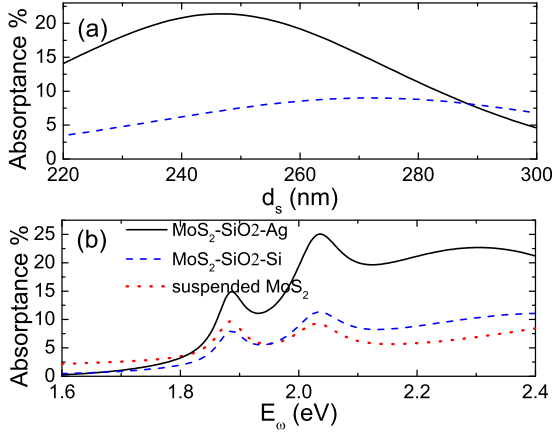}
\caption{(a) The absorptance of the monolayer MoS$_{2}$ with silver films (black solid line) and Si substrate (blue dashed line) as function of the spacer layers thickness $d_{s}$ for $E_{\omega}=2.21 eV$. (b) The absorptance of the monolayer MoS$_{2}$ as function of the photon energy for suspended monolayer MoS$_{2}$
(red dotted line), monolayer MoS$_{2}$  with Si substrate and 270 nm spacer layer (blue dashed line), and monolayer MoS$_{2}$  with sliver films and 247 nm spacer layer (black solid line).
}
\label{fig3}%
\end{figure}

Finally, we turn to the discussion on the absorptance of the monolayer MoS$_{2}$ prepared on top of Si substrate with SiO$_{2}$ spacer layer. This structure has been fabricated in the experiment \cite{13OLS}.  The Si substrate can also act as the RBR due to its high refractive index, and thus enhance the absorptance of the monolayer MoS$_{2}$.  The absorptance of the monolayer MoS$_{2}$ with  Si substrate as function of the spacer layers thickness $d_{s}$ for optical wavelength $\lambda=561 nm$ (same as in the experiment \cite{13OLS}). For $d_{s}=270$ nm, the maximum  absorptance of the monolayer MoS$_{2}$ is about  9\% [Fig. 3(a)]. This is why a $d_{s}=270$ nm SiO$_{2}$ spacer layer is fabricated in experiment. By contrast, the absorptance of suspended monolayer MoS$_{2}$ is about 5.9\%  and the absorptance of  monolayer MoS$_{2}$ with sliver films is about 22\% due to the higher reflectivity of sliver films. Furthermore, the FWHM of the absorption spectrum of monolayer MoS$_{2}$ with metal films is larger than that of monolayer MoS$_{2}$ with Si substrate [Fig. 3(b)]. Thus if the Si substrate is replaced by the metal films in the experiment,  the photoresponsivity and the FWHM of monolayer MoS$_{2}$ photodetectors  can be further enhanced greatly.

In conclusion, the optical absorption of monolayer MoS$_{2}$ prepared on top of 1DPC or silver films with a spacer layer is investigated theoretically. In these structures, the maximum optical absorptance of monolayer MoS$_{2}$ can be raised to 35\% with a large FWHM due to the F-P cavity effect. The absorption of monolayer MoS$_{2}$ with 1DPC is slightly larger than that of monolayer MoS$_{2}$ with silver films.  However, the  FWHM of the absorption spectrum of monolayer  MoS$_{2}$  with silver film is much larger than that of monolayer MoS$_{2}$ with 1DPC. The absorption of monolayer MoS$_{2}$ can also be tuned by varying the thickness of the spacer layers and cover layers. Our proposal is very easy to implement and may have potentially important  applications in the development of  monolayer MoS$_{2}$ based optoelectronic devices.

This work was supported by the NSFC Grant Nos. 11264029, 11264030, and  11364033, the NSF from the Jiangxi Province No. 20122BAB212003.

\section*{Added Materials}

In the front section, the  permittivity of monolayer MoS$_{2}$  is extracted from experiment results\cite{10KFM}.  In Ref. \cite{10KFM}, experimental measurement is in the range about 1.6-2.4 eV.  In the Fig. 2(d) in Ref. \cite{13DYQ}, a experiment result of the absorption monolayer MoS$_{2}$ in the range about 1.6-2.9 eV is present. Next, we extract The permittivity of monolaye MoS$_{2}$ in the range about 1.6-2.9 eV  from the Fig. 2(d) in Ref. \cite{13DYQ}. Similar Eq. (\ref{EQ_eps}) but  using two band transition, the imaginary part of the total permittivity can be expressed as
\begin{eqnarray}
\varepsilon _{i} &=&\frac{f_{ex}^{A}\Gamma _{A}}{(E_{\omega
}-E_{ex}^{A})^{2}+\Gamma _{A}^{2}}+\frac{f_{ex}^{B}\Gamma _{B}}{%
(E_{\omega }-E_{ex}^{B})^{2}+\Gamma _{B}^{2}} \nonumber \\
&+&\frac{f_{b}^{A}e}{\hbar\omega}\Theta (E_{\omega } -E_{g}^{A},\Gamma_{band}^{A}) \nonumber \\
&+&\frac{f_{b}^{B}e}{\hbar\omega}\Theta (E_{\omega } -E_{g}^{B},\Gamma_{band}^{B}), \label{EQ_eps2}
\end{eqnarray}
where  $f_{b}^{A}$ ($f_{b}^{B}$),  $E_{g}^{A}$ ($E_{g}^{B}$), and $\Gamma_{band}^{A}$ ($\Gamma_{band}^{B}$) are equivalent oscillator strength of interband transition, the linewidth, band gap, and linewidth of the interband transition of  A (B) band, respectively, $E_{g}^{B}=E_{g}^{A}+\Delta_{band}$,  $\Delta_{band}=146$ meV is used in the calculation. The real part of the permittivity of monolayer is obtained  by using Eq. (\ref{EQ_epsr}). $n_{r0}$ is adjusted to make $\varepsilon_{r}(\sim0)\sim 4.2$ (if the real part of the permittivity is not too large, the absorption of the  monolayer MoS$_{2}$ is only depended on the imaginary part of the  permittivity). To compare with the experiment results, the absorbance of monolayer MoS$_{2}$ is calculated by using the same method as in Refs. \cite{10KFM} and \cite{08KFM}.

 Results are shown in Fig. 4 and table 3 (at the last page  of the paper). Compared with results in table 1, a little changes of the parameters of A and B exicitons can be found. However, the parameters of the band transition fitted by Eq. (\ref{EQ_eps2}) is different with that fitted by using Eqs. (\ref{EQ_eps}).

\begin{figure}[t]
\centering
\includegraphics[width=0.9\columnwidth,clip]{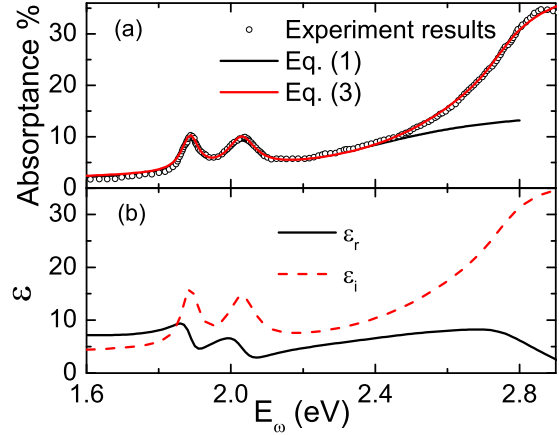}
\caption{(a) The absorbance of monolayer MoS$_{2}$ as a function of the light wavelength of the experimental results (black solid-circle curve)\cite{13DYQ},  fit results  by using Eq. (\ref{EQ_eps2}) (red solid line), and  fit results  by using Eq. (\ref{EQ_eps}) (black solid line). (b)Fitted permittivity of monolayer by using Eqs. (\ref{EQ_eps2}) and (\ref{EQ_epsr}).
}
\label{fig4}\label{fig4}
\end{figure}

\begin{table}[t]
  \caption{Fitted parameters in Eq. (\ref{EQ_eps2})}
  \begin{center}
    \begin{tabular}{ccccccccc}
    \hline
    \hline
    $E_{ex}^{A}$ & $\Gamma _{A}$ & $E_{ex}^{B}$ & $\Gamma _{B}$ \\
    \hline
    1.887 eV & 28 meV &  2.03 eV & 46 meV \\
    \hline
    \hline
    $f_{ex}^{A}$  &   $f_{ex}^{B}$ & $E_{g}^{A}$\\
    \hline
      0.29 eV & 0.40 eV & 2.60 eV  \\
    \hline
    \hline
         $f_{b}^{A}$ & $f_{b}^{B}$ & $\Gamma_{band}^{A}$ &  $\Gamma_{band}^{B}$\\
    \hline
       74 & 50 & 244 meV & 82 meV\\
    \hline
    \hline
    \end{tabular}
  \end{center}
\end{table}

\newpage
\begin{table}
\begin{center}
 \caption{Fitted permittivity of monolayer}
\begin{tabular}{Ic|c|cIc|c|cIc|c|cI}
   \whline
$E_{\omega}$ (eV)    &   $\varepsilon_{r}$ &   $\varepsilon_{i}$ &   $E_{\omega}$ (eV)    &   $\varepsilon_{r}$ &   $\varepsilon_{i}$ &   $E_{\omega}$ (eV)    &   $\varepsilon_{r}$ &   $\varepsilon_{i}$\\
\shline
  1.60  & 7.17  & 4.43  & 2.04  & 4.01  &14.29  & 2.48  & 7.26  &12.61 \\
  1.61  & 7.16  & 4.46  & 2.05  & 3.39  &13.32  & 2.49  & 7.34  &12.94 \\
  1.62  & 7.16  & 4.48  & 2.06  & 3.05  &12.14  & 2.50  & 7.42  &13.29 \\
  1.63  & 7.15  & 4.51  & 2.07  & 2.95  &11.03  & 2.51  & 7.49  &13.65 \\
  1.64  & 7.16  & 4.54  & 2.08  & 2.99  &10.11  & 2.52  & 7.56  &14.02 \\
  1.65  & 7.16  & 4.58  & 2.09  & 3.11  & 9.39  & 2.53  & 7.63  &14.40 \\
  1.66  & 7.17  & 4.61  & 2.10  & 3.27  & 8.85  & 2.54  & 7.69  &14.79 \\
  1.67  & 7.18  & 4.65  & 2.11  & 3.44  & 8.45  & 2.55  & 7.75  &15.20 \\
  1.68  & 7.20  & 4.70  & 2.12  & 3.61  & 8.15  & 2.56  & 7.81  &15.62 \\
  1.69  & 7.23  & 4.74  & 2.13  & 3.78  & 7.94  & 2.57  & 7.87  &16.06 \\
  1.70  & 7.26  & 4.79  & 2.14  & 3.94  & 7.79  & 2.58  & 7.92  &16.50 \\
  1.71  & 7.29  & 4.85  & 2.15  & 4.09  & 7.69  & 2.59  & 7.97  &16.96 \\
  1.72  & 7.33  & 4.91  & 2.16  & 4.23  & 7.62  & 2.60  & 8.01  &17.44 \\
  1.73  & 7.38  & 4.98  & 2.17  & 4.36  & 7.59  & 2.61  & 8.05  &17.92 \\
  1.74  & 7.44  & 5.06  & 2.18  & 4.48  & 7.58  & 2.62  & 8.09  &18.43 \\
  1.75  & 7.51  & 5.15  & 2.19  & 4.60  & 7.59  & 2.63  & 8.13  &18.95 \\
  1.76  & 7.59  & 5.26  & 2.20  & 4.72  & 7.62  & 2.64  & 8.16  &19.50 \\
  1.77  & 7.68  & 5.38  & 2.21  & 4.83  & 7.66  & 2.65  & 8.19  &20.07 \\
  1.78  & 7.79  & 5.53  & 2.22  & 4.93  & 7.72  & 2.66  & 8.22  &20.66 \\
  1.79  & 7.92  & 5.72  & 2.23  & 5.04  & 7.79  & 2.67  & 8.24  &21.29 \\
  1.80  & 8.07  & 5.95  & 2.24  & 5.14  & 7.87  & 2.68  & 8.25  &21.95 \\
  1.81  & 8.25  & 6.24  & 2.25  & 5.23  & 7.96  & 2.69  & 8.25  &22.65 \\
  1.82  & 8.46  & 6.63  & 2.26  & 5.33  & 8.06  & 2.70  & 8.23  &23.40 \\
  1.83  & 8.70  & 7.18  & 2.27  & 5.42  & 8.17  & 2.71  & 8.19  &24.18 \\
  1.84  & 8.97  & 7.95  & 2.28  & 5.52  & 8.28  & 2.72  & 8.12  &25.01 \\
  1.85  & 9.23  & 9.09  & 2.29  & 5.61  & 8.41  & 2.73  & 8.02  &25.86 \\
  1.86  & 9.38  &10.78  & 2.30  & 5.70  & 8.54  & 2.74  & 7.87  &26.73 \\
  1.87  & 9.16  &13.09  & 2.31  & 5.79  & 8.69  & 2.75  & 7.68  &27.60 \\
  1.88  & 8.15  &15.39  & 2.32  & 5.88  & 8.84  & 2.76  & 7.44  &28.44 \\
  1.89  & 6.45  &16.02  & 2.33  & 5.97  & 9.00  & 2.77  & 7.17  &29.24 \\
  1.90  & 5.09  &14.46  & 2.34  & 6.06  & 9.17  & 2.78  & 6.86  &29.98 \\
  1.91  & 4.61  &12.30  & 2.35  & 6.15  & 9.35  & 2.79  & 6.53  &30.66 \\
  1.92  & 4.68  &10.68  & 2.36  & 6.24  & 9.54  & 2.80  & 6.17  &31.27 \\
  1.93  & 4.97  & 9.69  & 2.37  & 6.33  & 9.73  & 2.81  & 5.81  &31.81 \\
  1.94  & 5.31  & 9.19  & 2.38  & 6.42  & 9.94  & 2.82  & 5.44  &32.30 \\
  1.95  & 5.65  & 9.04  & 2.39  & 6.51  &10.16  & 2.83  & 5.07  &32.72 \\
  1.96  & 5.97  & 9.17  & 2.40  & 6.59  &10.38  & 2.84  & 4.71  &33.10 \\
  1.97  & 6.24  & 9.55  & 2.41  & 6.68  &10.62  & 2.85  & 4.34  &33.43 \\
  1.98  & 6.46  &10.16  & 2.42  & 6.77  &10.87  & 2.86  & 3.98  &33.71 \\
  1.99  & 6.58  &11.03  & 2.43  & 6.85  &11.13  & 2.87  & 3.63  &33.97 \\
  2.00  & 6.54  &12.12  & 2.44  & 6.94  &11.40  & 2.88  & 3.29  &34.19 \\
  2.01  & 6.24  &13.30  & 2.45  & 7.02  &11.69  & 2.89  & 2.95  &34.39 \\
  2.02  & 5.65  &14.27  & 2.46  & 7.10  &11.98  & 2.90  & 2.62  &34.56 \\
  2.03  & 4.83  &14.67  & 2.47  & 7.19  &12.29  &  &   &   \\

  \whline
\end{tabular}
\end{center}
\end{table}

%\bibliography{mybib}

\begin{thebibliography}{10}
\newcommand{\enquote}[1]{``#1''}

\bibitem{12QHW}
Q.~H. Wang, K.~Kalantar-Zadeh, A.~Kis, J.~N. Coleman, and M.~S. Strano, Nat.
  Nanotech. \textbf{7}, 699 (2012).

\bibitem{10KFM}
K.~F. Mak, C.~Lee, J.~Hone, J.~Shan, and T.~F. Heinz, Phys. Rev. Lett.
  \textbf{105}, 136805 (2010).

\bibitem{10AS}
A.~Splendiani, L.~Sun, Y.~Zhang, T.~Li, J.~Kim, C.~Y. Chim, G.~Galli, and
  F.~Wang, Nano Lett. \textbf{10}, 1271 (2010).

\bibitem{07TL}
T.~Li and G.~Galli, J. Phys. Chem. C \textbf{111}, 16192 (2007).

\bibitem{13OLS}
O.~Lopez-Sanchez, D.~Lembke, M.~Kayci, A.~Radenovic, and A.~Kis, Nat. Nanotech.
  \textbf{8}, 497 (2013).

\bibitem{11ZY}
Z.~Yin, H.~Li, H.~Li, L.~Jiang, Y.~Shi, Y.~Sun, G.~Lu, Q.~Zhang, X.~Chen, and
  H.~Zhang, ACS Nano \textbf{6}, 74 (2011).

\bibitem{12HSL}
H.~S. Lee, S.~W. Min, Y.~G. Chang, M.~K. Park, T.~Nam, H.~Kim, J.~H. Kim,
  S.~Ryu, and S.~Im, Nano Lett. \textbf{12}, 3695 (2012).

\bibitem{12WC}
W.~Choi, M.~Y. Cho, A.~Konar, J.~H. Lee, G.~Cha, S.~C. Hong, S.~Kim, J.~Kim,
  D.~Jena, J.~Joo, and S.~Kim, Adv. Mater. \textbf{24}, 5832 (2012).

\bibitem{13MB}
M.~Bernardi, M.~Palummo, and J.~C. Grossman, Nano Lett. \textbf{13}, 3664
  (2013).

\bibitem{12JP}
J.~Pu, Y.~Yomogida, K.~K. Liu, L.~J. Li, Y.~Iwasa, and T.~Takenobu, Nano Lett.
  \textbf{12}, 4013 (2012).

\bibitem{14CJ}
C.~Janisch, N.~Mehta, D.~Ma, A.~L. El¨ªas, N.~Perea-L¨®pez, M.~Terrones, and
  Z.~Liu, Opt. Lett. \textbf{39}, 383 (2014).

\bibitem{14AS}
A.~Sobhani, A.~Lauchner, S.~Najmaei, C.~Ayala-Orozco, F.~Wen, J.~Lou, and N.~J.
  Halas, Appl. Phys. Lett. \textbf{104}, 031112 (2014).

\bibitem{13XG}
X.~Gan, Y.~Gao, K.~F. Mak, X.~Yao, R.~J. Shiue, A.~van~der Zande, M.~E.
  Trusheim, F.~Hatami, T.~F. Heinz, J.~Hone, and D.~Englund, Appl. Phys. Lett.
  \textbf{103}, 181119 (2013).

\bibitem{12YVB}
Y.~V. Bludov, M.~I. Vasilevskiy, and N.~M.~R. Peres, J. Appl. Phys.
  \textbf{112}, 084320 (2012).

\bibitem{12ST}
S.~Thongrattanasiri, F.~H.~L. Koppens, and F.~J.~G. de~Abajo, Phys. Rev. Lett.
  \textbf{108}, 047401 (2012).

\bibitem{12AFa}
A.~Ferreira, N.~M.~R. Peres, R.~M. Ribeiro, and T.~Stauber, Phys. Rev. B
  \textbf{85}, 115438 (2012).

\bibitem{12MAV}
M.~A. Vincenti, D.~de~Ceglia, M.~Grande, A.~D'Orazio, and M.~Scalora, Opt.
  Lett. \textbf{38}, 3550 (2013).

\bibitem{12AFb}
A.~Ferreira and N.~M.~R. Peres, Phys. Rev. B \textbf{86}, 205401 (2012).

\bibitem{08ZZZ}
Z.~Z. Zhang, K.~Chang, and F.~M. Peeters, Phys. Rev. B \textbf{77}, 235411
  (2008).

\bibitem{13WZ}
W.~Zhao, K.~Shi, and Z.~Lu, Opt. Lett. \textbf{38}, 4342 (2013).

\bibitem{13QY}
Q.~Ye, J.~Wang, Z.~Liu, Z.~C. Deng, X.~T. Kong, F.~Xing, X.~D. Chen, W.~Y.
  Zhou, C.~P. Zhang, and J.~G. Tian, Appl. Phys. Lett \textbf{102}, 021912
  (2013).

\bibitem{12RA}
R.~Alaee, M.~Farhat, C.~Rockstuhl, and F.~Lederer, Optics Express \textbf{20},
  28017 (2012).

\bibitem{13XZ}
X.~Zhu, W.~Yan, P.~U. Jepsen, O.~Hansen, N.~A. Mortensen, and S.~Xiao, Appl.
  Phys. Lett. \textbf{102}, 131101 (2013).

\bibitem{12JTL}
J.~T. Liu, N.~H. Liu, J.~Li, X.~J. Li, and J.~H. Huang, Appl. Phys. Lett
  \textbf{101}, 052104 (2012).

\bibitem{13NMRP}
N.~M.~R. Peres and Y.~V. Bludov, EPL \textbf{101}, 58002 (2013).

\bibitem{13JTL}
J.~T. Liu, N.~H. Liu, L.~Wang, X.~H. Deng, and F.~H. Su, EPL \textbf{104},
  57002 (2013).

\bibitem{85EDP}
E.~D. Palik, ed., \emph{Handbook of Optical Constants of Solids} (Academic
  Press, Boston, 1985).

\bibitem{13DYQ}
D.~Y. Qiu, F.~H. da~Jornada, and S.~G. Louie, Phys. Rev. Lett. \textbf{111},
  216805 (2013).

\bibitem{13HS}
H.~Shi, H.~Pan, Y.~W. Zhang, and B.~I. Yakobson, Phys. Rev. B \textbf{87},
  155304 (2013).

\bibitem{12TC}
T.~Cheiwchanchamnangij and W.~R.~L. Lambrecht, Phys. Rev. B \textbf{85}, 205302
  (2012).

\bibitem{12AR}
A.~Ramasubramaniam, Phys. Rev. B \textbf{86}, 115409 (2012).

\bibitem{11BR}
B.~Radisavljevic, A.~Radenovic, J.~Brivio, V.~Giacometti, and A.~Kis, Nat. Nanotech. \textbf{6}, 147 (2011).

\bibitem{11YY}
Y.~Yoon, K.~Ganapathi, and S.~Salahuddin, Nano Lett. \textbf{11}, 3768 (2011).

\bibitem{08KFM}
Kin Fai Mak, Matthew Y. Sfeir, Yang Wu, Chun Hung Lui, James A. Misewich, and Tony F. Heinz, Phys. Rev. Lett. \textbf{101}, 196405 (2008).

\end{thebibliography}
%\bibliographystyle{ol}

\end{document}